\documentclass[aps,twocolumn,amsmath,amssymb,showpacs,prb,superscriptaddress,unsortedaddress]{revtex4}
\usepackage{epsf}
\usepackage{graphicx}

\begin{document}

\title{Evidence for two energy scales in the superconducting state of\\ optimally doped (Bi,Pb)$_2$(Sr,La)$_2$CuO$_{6+\delta}$}

\author{Takeshi Kondo}
\affiliation{Department of Crystalline Materials Science, Nagoya University, 
             Nagoya 464-8603, Japan}
\affiliation{Ames Laboratory and Department of Physics and Astronomy, Iowa State University, Ames, IA 50011, USA}

\author{Tsunehiro Takeuchi}
\affiliation{Department of Crystalline Materials Science, Nagoya University, 
             Nagoya 464-8603, Japan}
\affiliation{EcoTopia Science Institute, Nagoya University, Nagoya 464-8603, Japan}

\author{Adam Kaminski}
\affiliation{Ames Laboratory and Department of Physics and Astronomy, Iowa State University, Ames, IA 50011, USA}

\author{Syunsuke Tsuda}
\affiliation{Institute of Solid State Physics, University of Tokyo, Kashiwa 277-8581, Japan}

\author{Shik Shin}
\affiliation{Institute of Solid State Physics, University of Tokyo, Kashiwa 277-8581, Japan}

\date{\today}
\begin{abstract}
We use angle-resolved photoemission spectroscopy (ARPES) to investigate the properties of the energy gap(s) in optimally doped (Bi,Pb)$_2$(Sr,La)$_2$CuO$_{6+\delta}$ (Bi2201). We find that the spectral gap has two components  in the superconducting state: a superconducting gap and pseudogap. Significant differences in their momentum and temperature dependence suggest that they represent two separate energy scales. Spectra near the node reveal a sharp peak with a small gap below $T_c$ that closes at $T_c$. Near the antinode, the spectra are broad with a large energy gap of $\sim$40meV above and below $T_c$. The spectral shape and gap magnitude around the antinode are almost constant across $T_c$, which indicates that the pseudogap state coexists with the superconducting state below $T_c$, and it dominates the character of the spectra around the antinode. We speculate that the pseudogap state competes with the superconductivity by diminishing spectral weight in the antinodal regions, where the magnitude of the superconducting gap is largest.\end{abstract}


\maketitle
The pseudogap is one of the most fascinating properties of the high temperature superconductors\cite{Timusk}.  It gives rise to a strange state of matter above $T_c$ where parts of the Fermi surface consist of disconnected "arcs" \cite{Norman_nature}, while the remainder is gapped.
Recent Angle Resolved Photoemission Spectroscopy (ARPES) measurements show that the pseudogap state extrapolates at absolute zero to a nodal liquid where the Fermi surface consists of just four points\cite{Kanigel}. Since the pseudogap is often closely linked to the mechanism of high temperature superconductivity, it is very important to understand its properties and relationship to the superconducting gap.  
According to one class of theories\cite{Emery}, the pseudogap opens because electrons are paired at temperatures much higher than the critical temperature ($T_c$) with the same pairing mechanism as the superconducting gap. 
The pairs only condense when the sample is cooled to $T_c$.
This scenario is supported by a number of ARPES studies on mostly Bi$_2$Sr$_2$CaC$_2$O$_{8+\delta}$ (Bi2212) suggesting that the behavior and symmetry of the pseudogap  above $T_c$ are similar to those of the superconducting gap below $T_c$\cite{DING_PG,LOESER_PG,JC_SCALING}. 
Another class of theories\cite{AFFLECK,Varma97,Varma99}  link the pseudogap to an ordered state with possibly a separate energy scale. 
The first ARPES experiment designed to detect an ordered state below the pseudogap temperature gave a positive result\cite{KAMINSKITRSB}. However, a later ARPES experiment was unable to detect the same small signatures  in the data\cite{BORISENKOTRSB}. 
More recently, a high precision neutron scattering experiment provided direct evidence for the existence of an ordered state of particular symmetry below the pseudogap temperature\cite{MagneticOrder}. This result confirmed the predictions of Varma and was in agreement with the first ARPES study. 
Recent scanning tunneling microscopy and scanning tunneling spectroscopy (STM/STS) experiments on Bi2212 show that even below $T_c$, a pseudogap state, characterized by a large gap and broad spectral peaks, coexists with the superconducting state, which has a smaller energy gap and sharp spectral peaks\cite{Pan,Mcelroy}.  
One drawback of studying the pseudogap in Bi2212 is its very large superconducting gap ($\sim$40 meV at optimal doping), which is comparable to the pseudogap. We chose to study (Bi,Pb)$_2$(Sr,La)$_2$CuO$_{6+\delta}$ (Bi2201), which has a low $T_c$ of $\sim$35K at optimal doping. NMR \cite{Zheng} and electrical resistivity \cite{Ando} experiments estimate the pseudogap temperature in Bi2201 to be similar to that of Bi2212, while $T_c$ is almost three times smaller. 
Therefore, we should gain an important insight into the relationship between the pseudogap and the superconducting gap by directly measuring the energy gap in Bi2201. 
In this letter, we report ARPES measurements of the momentum- and temperature-dependence of the energy gap in optimally doped Bi2201 with $T_c=35$K.  
On first inspection, the momentum dependence of the energy gap below $T_c$ strongly deviates from the symmetry of a monotonic $d_{x^2  - y^2 }$ wave function, which is observed in optimally doped Bi2212\cite{Ding_gap}. 
The gap symmetry and its temperature dependence are most consistent with a two gap component model: a small $d_{x^2  - y^2 }$ wave superconducting gap that dominates the symmetry near the node and a large pseudogap that exists only around the antinode.  

We measured optimally doped single crystals  of (Bi,Pb)$_2$(Sr,La)$_2$CuO$_{6+\delta}$ ($T_c=$35K) that were grown using a conventional floating-zone (FZ) technique and subsequent annealing\cite{Kondo}. $T_c$ was determined from susceptibility and electrical resistivity measurements and the samples had a sharp superconducting transition ($\sim$3meV). Pb was used to substitute for Bi in order to suppress the modulation of the BiO plane. This normally leads to a contamination of the ARPES signal from umklapp replicas of the main band. The use of modulation-free samples enabled us to precisely determine the energy gap. The measurements were made using a Scienta SES2002 hemispherical analyzer with a Gammadata VUV5010 photon source (HeI$\alpha$) at the Institute of Solid State Physics (ISSP), the University of Tokyo. The energy and angular resolutions were 5meV and $0.13 ^\circ$, respectively.

\begin{figure}
\includegraphics[width=2.8in]{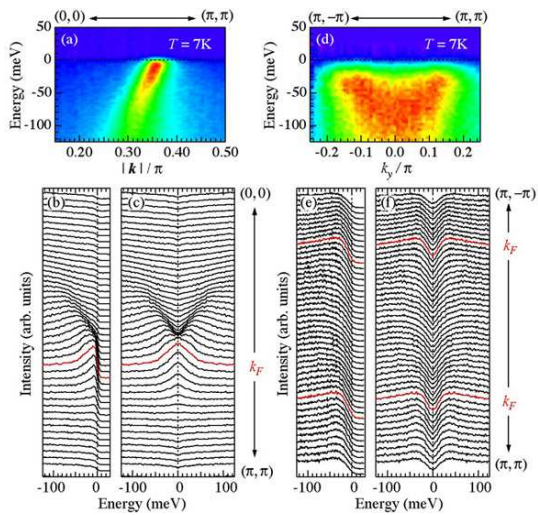}
\caption{(color online) ARPES intensity and the corresponding EDCs measured at $7$K well below $T_c$ along (a,b) (0,0)-($\pi,\pi$)  and (d,e) ($\pi,-\pi$)-($\pi,\pi$) cut. (c,f) Symmetrized EDC of (b,c).}
\label{fig1}
\end{figure}

In Fig. 1 (a,b) and (d,e), we show the ARPES data (intensity plots and the corresponding Energy Distribution Curves, EDCs) measured at 7K well below $T_c$ in the nodal and antinodal regions, respectively.  The nodal spectra are characterized by sharp peaks. The leading edge of the EDC at the nodal $\mbox{\boldmath $k$}_F$ reaches the Fermi level, indicating there is no energy gap. The spectra at the antinodal cut, in contrast, are very broad. EDCs near the antinodal $\mbox{\boldmath $k$}_F$ are shifted towards higher binding energies due to the presence of an energy gap. This effect is illustrated more easily in Fig. 1 (c) and (f) by use of a symmetrization method \cite{Norman_nature}. EDCs are reflected about the Fermi level and added to the original data. This technique removes the Fermi function and enables us to immediately identify the presence of an energy gap. Clearly, a gap is present in the data of Fig. 1 (f) and is absent in Fig. 1 (c). 

\begin{figure}
\includegraphics[width=3.4in]{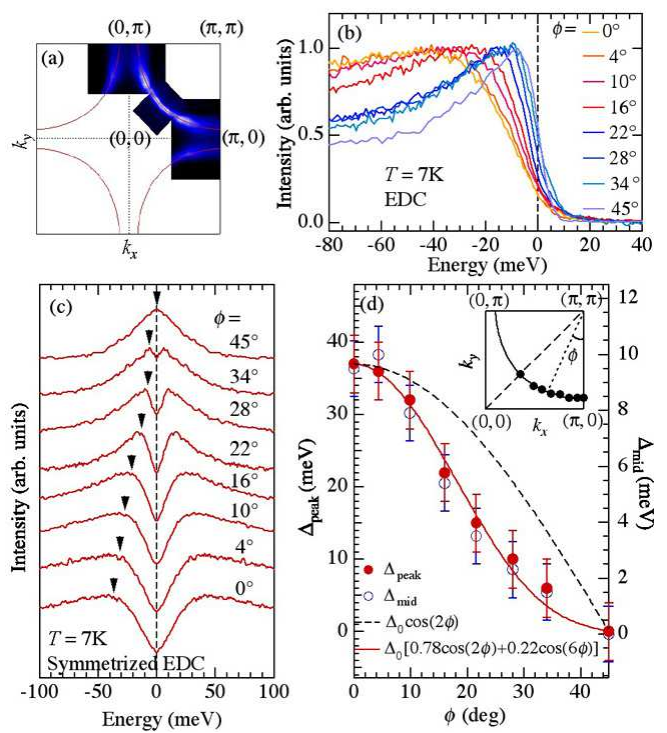}
\caption{(color online) (a) ARPES intensity map around the Fermi level as a function of $k_x$ and $k_y$.(b) EDC and (c) symmetrized EDC measured at  $T=7$K (well below $T_c$) for various $\mbox{\boldmath $k$}_F$. The angle $\phi$ is defined in the inset of (d). (d) Magnitude of the energy gap around the Fermi surface at $T=7$K estimated from the mid-point of the EDC leading edge ($\Delta _{{\rm{mid}}}$) and the peak position in the symmetrized EDC ($\Delta _{{\rm{peak}}}$ - indicated by the arrows in (c)). The solid line shows a gap function of  the form $\Delta (\phi ) = \Delta _0 [0.78\cos (2\phi ) + 0.22\cos (6\phi )]$ (dashed line corresponds to $\Delta (\phi ) = \Delta _0 \cos (2\phi )$). }
\label{fig2}
\end{figure}

Figure. 2(a) shows the ARPES intensity at the Fermi level. The intensity is strongest near the node and diminishes towards the antinode ($\pi$,0). This is a clear signature of an anisotropic energy gap.
We determined the size of the energy gap using two methods. The first estimates the shift in energy of the mid point  of the EDC leading edge ($\Delta _{{\rm{mid}}} $).  Figure 2(b) shows the EDCs measured at different $\mbox{\boldmath $k$}_F$ for angles $\phi$, defined in the inset of Fig. 2(d). The leading edge shift is zero at the node ($\phi=45^\circ$), and it increases toward the antinode ($\phi=0^\circ$). 
The second method determines the peak position of the symmetrized EDC ($\Delta _{{\rm{peak}}}$). Figure 2(c) shows results from this method with the peak positions marked by arrows.  Results from both methods are compared in Fig. 2(d).
It is clear that both methods show the gap increases towards the antinode. Although the maximum values are different for obvious reasons, both yield a similar symmetry of the gap. 
A remarkable feature in the energy gap is the strong deviation from a monotonic $d_{x^2-y^2}$-wave symmetry, which is illustrated by the dashed line in Fig. 2(d). 
Harris $et\ al$.\cite{Harris} first reported this feature in Bi2201 about a decade ago, and suggested that the enhanced anisotropy is caused by impurity scattering which results in a strong suppression of the energy gap close to the node.
The authors measured samples with a large residual resistivity ratio, $\rho _{ab}$(300K)/$\rho _{ab} $(0K), of 2.4, so idea of an impurity-induced gap symmetry seemed plausible\cite{Haas}.
(Here, $\rho _{ab}$ represents the resistivity along the CuO$_2$ plane, and $\rho _{ab} $(0K) was estimated from an interpolation to $T=0$K.) 
In this work, we employed high quality single crystals with a low $\rho _{ab}$(300K)/$\rho _{ab} $(0K) ratio of $\sim7$, yet we observe a similar deviation from the monotonic $d_{x^2-y^2}$-wave symmetry. Hence, impurities are unlikely to be the cause of the characteristic gap symmetry in Bi2201.

It has been reported that the superconducting gap also significantly deviates from a monotonic $d_{x^2  - y^2 }$-wave symmetry in underdoped Bi2212\cite{Mesot}. This was attributed to an increase of the electron correlation with underdoping, which may increase the intensity of the higher order harmonic component in the $d$-wave gap function. 
We fitted a function of the form $\Delta (\phi ) = \Delta _0 [B\cos (2\phi ) + (1 - B)\cos (6\phi )]$ to the energy gap in the present data, with a cos($6\phi$) second order harmonics term in the $d_{x^2  - y^2 }$-wave. The result is plotted in Fig. 2(d) using a solid line.
Even though our Bi2201 samples are optimally doped we find a much stronger deviation from a pure  $d_{x^2  - y^2 }$ symmetry ($B=0.78$) compared to underdoped Bi2212 with $T_c=75$K ($B=0.88$)\cite{Mesot}. Hence, the deviation in Bi2201 is unlikely to be due to strong correlation effects.

\begin{figure}
\includegraphics[width=3.4in]{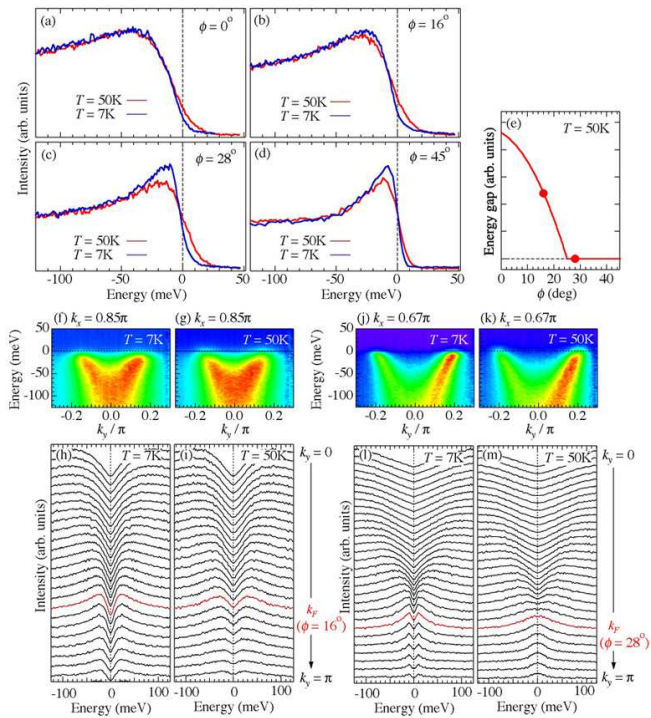}
\caption{(color online) (a,b,c,d) EDC below and above $T_c$ (7K and 50K) at various $\mbox{\boldmath $k$}_F$'s ($\phi=0^\circ, 16^\circ, 28^\circ$, and $45^\circ$). (e) Schematic illustration of the energy gap above $T_c$ as a function of the angle $\phi$. The two solid circles correspond to the energy gap at $\phi=16^\circ$ and $28^\circ$.
ARPES intensity map and the corresponding symmetrized EDCs measured on a momentum cut at $k_x=0.85\pi$ crossing $\mbox{\boldmath $k$}_F$ of $\phi=16^\circ$ (f,g,h,i) and at $k_x=0.65\pi$ crossing $\mbox{\boldmath $k$}_F$ of $\phi=28^\circ$ (j,k,l,m). (The contrast of the ARPES intensity between the positive and negative $k_y$ is caused by the matrix element effect.) Left and right hand panels were obtained at $T=7$K and 50K, respectively.}
\label{fig3}
\end{figure}

Figure 3 (a)-(d) show EDCs at four $\mbox{\boldmath $k$}_F$s above and below $T_c$ (7K and 50K). 
Spectra near the node (Fig.3 (c) and (d)) vary significantly across $T_c$; the sharp peak below $T_c$ broadens above $T_c$ and the energy gap closes slightly away from the node. 
This is contrasted by very little variation in the spectral line shape near the antinode across $T_c$ (Fig.3 (a) and (b)). The main feature here is a slight increase in spectral intensity above $T_c$ very close to the Fermi level. The two extreme behaviors evolve rapidly with momentum as evident by comparing panels  (f-i) and (j-m). They show the ARPES intensity and symmetrized EDCs measured along two momentum cuts crossing $\mbox{\boldmath $k$}_F$ at angles $\phi=16^\circ$ and $28^\circ$, respectively. Above $T_c$, the gap closes at $\mbox{\boldmath $k$}_F$ corresponding to $\phi=28^\circ$. For $\phi=16^\circ$ the symmetrized EDC shows a weak variation in the spectral line shape and has an almost unchanged peak position. This generates a large Fermi arc above $T_c$ schematically illustrated in Fig. 3 (e).
Previous results from Bi2212\cite{Norman_nature,DING_PG} reported a small spectral variation with increasing temperature in the pseudogap state above $T_c$. Therefore our current data on Bi2201 indicates that the energy gap near the antinode is dominated by the pseudogap even below $T_c$.

In order to investigate the pseudogap state, we measured ARPES spectra over a wide temperature range.  
The resulting symmetrized EDCs are superimposed in Fig. 4(a)-(d) for several $\mbox{\boldmath $k$}_F$s. 
The pseudogap is quite large up to 100K and closes at $\sim$150K. In optimally doped Bi2212, the pseudogap closing temperature ($T_{PG}$) has been estimated to be $\sim$130K by ARPES\cite{Sato_OP}. Thus the present results indicate that the pseudogap (characterized by the energy gap size and $T_{PG}$) does not scale with $T_c$ in optimally doped high-$T_c$ cuprates. In Fig. 4 (i), we plot the measured energy gap ($\Delta _{{\rm{peak}}}$) as a function of  angle $\phi$  at several temperatures ranging from below $T_c$ to above $T_{PG}$. We also plot the monotonic $d_{x^2  - y^2 }$-wave function for the superconducting gap of optimally doped Bi2212\cite{Ding_gap,Mesot} (solid black line) and that scaled by the ratio of the $T_c$s in optimally doped Bi2201 and Bi2212 (dashed line). Around the node, the gap symmetry at 7K is consistent with the $d_{x^2  - y^2 }$-wave gap function (dashed line). This component disappears above $T_c$. The energy gap rapidly increases towards the antinode and its maximum value is very similar to that in optimally doped Bi2212. The characteristic gap symmetry below $T_c$ in optimally doped Bi2201, therefore, can be understood as a coexistence of the superconducting state with a small gap ($\sim$15 meV) that has a monotonic $d_{x^2  - y^2 }$-wave symmetry and a pseudogap state that has a large energy gap similar to that of optimally doped Bi2212. The former dominates the spectral lineshape around the node, while the latter dominates at the antinodal region. In Bi2212, the superconducting gap has a similar energy size as the pseudogap ($\sim$40meV at the antinode), thus these two different gaps are continuously connected across $T_c$ and appear to have the same origin\cite{DING_PG,LOESER_PG}. In Bi2201 the superconducting gap is much smaller, due to its low $T_c$, whereas the pseudogap remains large. The very different properties of these two gaps around the Fermi surface lead us to conclude there is no direct relationship between the pseudogap and superconducting gap. 

\begin{figure}
\includegraphics[width=3.5in]{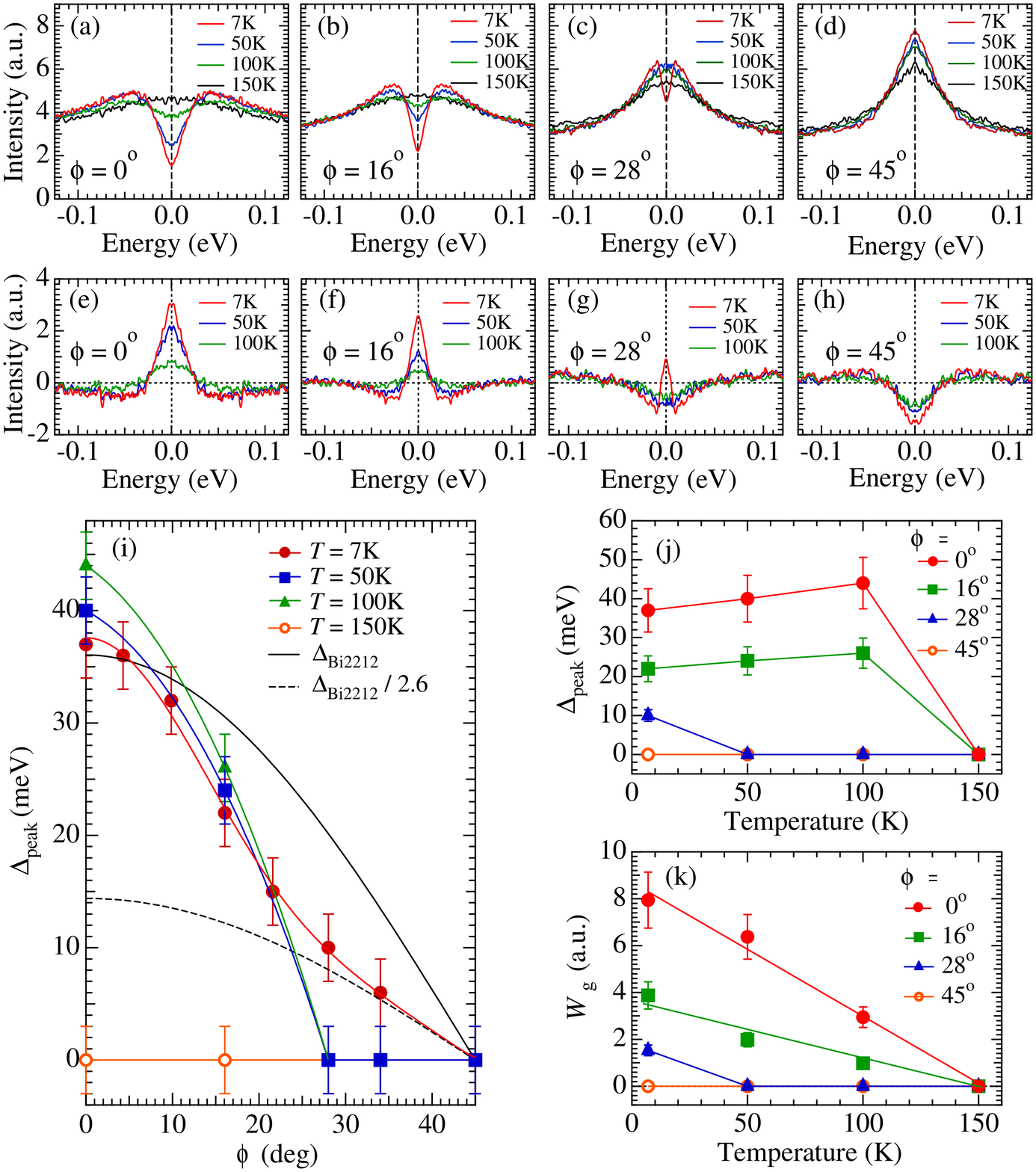}
\caption{(color online) (a,b,c,d) Symmetrized EDCs at various $\mbox{\boldmath $k$}_F$ ($\phi=0^\circ, 16^\circ, 28^\circ$, and $45^\circ$) and temperatures. (e,f,g,h) The symmetrized EDC at $T=$150K (above the pseudogap closing temperature, $T_{PG}$) subtracted by one at 7K, 50K, and 100K. (i) $\phi$ dependence of $\Delta_{\rm{peak}}$ from below $T_c$ to above $T_{PG}$. The solid black line shows $\Delta_{\rm{peak}}$ for optimally doped Bi2212 ($\Delta_{\rm{Bi2212}} \propto \cos (2\phi)$). The dashed black line shows $\Delta_{\rm{Bi2212}}$ divided by 2.6 ($\approx$$T_c$(Bi2212)/$T_c$(Bi2201)=$90{\rm{K}}/35{\rm{K}}$). (j) Peak position of the symmetrized EDC ($\Delta_{\rm{peak}}$) as a function of temperature. (k) Spectral weight lost due to the gap-opening near the Fermi energy ($W_{G}$) as a function of the temperature. }
\label{fig4}
\end{figure}

Finally, we should comment that while the spectral peak position is a good way to investigate the momentum behavior of the energy gap in the pseudogap state (Fig. 4 (i)), this is not true for its temperature dependence. We found that, similar to Bi2212\cite{Norman_nature,Norman_gapsize}, the peak position of the symmetrized EDCs around the antinode increases with increasing temperature below the pseudogap closing temperature ($T_{PG}$) and suddenly jump to zero above $T_{PG}$ as illustrated in Fig. 4(j).  This behavior should be contrasted with the continuous temperature dependence of the symmetrized spectral shape below $T_{PG}$ (Fig. 4 (a-d)). The significant temperature dependence of the pseudogap is illustrated by subtracting the symmetrized EDC below $T_{PG}$ from that above $T_{PG}$ as shown in Fig. 4(e-h). Here we note that when the energy gap is zero as shown at the node (Fig. 4(d)), the difference spectrum has a dip (Fig. 4(h)) because it simply reflects the thermal broadening of the spectral peak.  The difference spectrum has a peak when the energy gap is finite, reflecting the loss of spectral weight due to the energy gap. We estimated the spectral weight lost when the gap opens ($W_{G}$) from the area of the spectral peak in Fig. 4(e-h), and plot it in Fig. 4 (k). Near the node, $W_{PG}$ is zero above $T_c$ because the superconducting gap closes. In contrast, $W_{G}$ around the antinode decreases with increasing temperature in an almost linear fashion up to $T_{PG}$. The behavior of $W_{PG}$  without a jump across $T_c$ supports the idea that the ARPES spectrum around the antinode is dominated by the pseudogap state even below $T_c$.

In conclusion we report the momentum and temperature dependence of the energy gap measured by ARPES in an optimally doped single layer cuprate Bi2201 with $T_c$ = 35K. While the superconducting gap, which closes at $T_c$, is observed around the node, the ARPES spectra is dominated by the pseudogap state below $T_c$ around the antinode, where the spectral weight close to the Fermi level fills in with increasing temperature with a $T$-linear behavior.  
Significant differences in the momentum and temperature dependence of the pseudogap and superconducting gap force us to conclude there is no direct relationship between the two gaps. This points to the possibility of a competition between the pseudogap and superconducting gap. 

We thank M. R. Norman, H. M. Fretwell and C. M. Varma for useful remarks. This work was supported by Director Office for Basic Energy Sciences, US DOE. The Ames Laboratory is operated for the US DOE by Iowa State University under Contract No. W-7405-ENG-82. 

{\it Note added:}  After completion of this work, we became aware of related work by K. Tanaka {\it et al.}, Science Express: 10.1126/science.1133411 that reached similar conclusions using data from underdoped Bi2212 samples.


\begin{thebibliography}{99}

\bibitem{Timusk}
T. Timusk and B. Statt, 
Rep. Prog. Phys. {\bf 62}, 61 (1999).

\bibitem{Norman_nature}
M. R. Norman $et\ al$., 
Nature {\bf 392}, 157 (1998).

\bibitem{Kanigel}
A. Kanigel $et\ al$., 
Nature physics {\bf 2}, 447 (2006).

\bibitem{Emery}
V. J. Emery $et\ al$., 
Nature {\bf 374}, 434 (1995).

\bibitem{DING_PG}
H. Ding $et\ al$., 
Nature {\bf 382}, 51 (1996).

\bibitem{LOESER_PG}
A. G. Loeser $et\ al$.,
Science {\bf 273}, 325 (1996).

\bibitem{JC_SCALING}
J. C. Campuzano $et\ al$., 
Phys. Rev. Lett. {\bf 83}, 3709 (1999).

\bibitem{AFFLECK}
T. C. Hsu, J. B. Marston, and I. Affleck, 
Phys. Rev. B. {\bf 43}, 2866 (1991).

\bibitem{Varma97}
C. M. Varma, 
Phys. Rev. B. {\bf 55}, 14554 (1997).

\bibitem{Varma99}
C. M. Varma, 
Phys. Rev. Lett. {\bf 83}, 3538 (1999).

\bibitem{KAMINSKITRSB}
A. Kaminski $et\ al$., 
Nature {\bf 416}, 610 (2002).

\bibitem{BORISENKOTRSB}
S. V. Borisenko $et\ al$., 
Phys. Rev. Lett. {\bf 92}, 207001 (2004).

\bibitem{MagneticOrder}
B. Fauqu\'e $et\ al$., 
Phys. Rev. Lett. {\bf 96}, 197001 (2006).

\bibitem{Pan}
S. H. Pan $et\ al$., 
Nature {\bf 413}, 282 (2001).

\bibitem{Mcelroy}
K. McElroy $et\ al$., 
Phys. Rev. Lett. {\bf 94}, 197005 (2005).

\bibitem{Zheng}
Guo-qing Zheng $et\ al$., 
Phys. Rev. Lett. {\bf 94}, 047006 (2005).

\bibitem{Ando}
Yoichi Ando $et\ al$., 
Phys. Rev. Lett. {\bf 93}, 267001 (2004).

\bibitem{Ding_gap}
H. Ding $et\ al$., 
Phys. Rev. B {\bf 54}, R9678 (1996).

\bibitem{Kondo}
T. Kondo $et\ al$., 
Phys. Rev. B {\bf 72}, 024533 (2005).

\bibitem{Harris}
J. M. Harris $et\ al$., 
Phys. Rev. Lett. {\bf 79}, 143 (1997).

\bibitem{Haas}
S. Haas $et\ al$., 
Phys. Rev. B {\bf 56}, 5108 (1997).

\bibitem{Mesot}
J. Mesot $et\ al$., 
Phys. Rev. Lett. {\bf 83}, 840 (1999).

\bibitem{Sato_OP}
T. Sato $et\ al$., 
Physica C {\bf 341-348}, 815 (2000).

\bibitem{Norman_gapsize}
M. R. Norman $et\ al$., 
Phys. Rev. B {\bf 57}, R11093 (1998).

\end{thebibliography}
\end{document}